\documentclass[twocolumn,showpacs,amsmath,amssymb,prd]{revtex4}

\usepackage{graphicx}
\usepackage{dcolumn}
\usepackage{bm}
\usepackage{epsfig}

\begin{document}


\title{The vicissitudes of ``cannonballs": a response to criticisms
by A.M.~Hillas \\
and a brief review of our claims
}

\author{Arnon Dar}%
\affiliation{Department of Physics and Space Research 
Institute, Technion, Haifa 32000, Israel}%

\author{A. De R\'ujula}
\affiliation{Theory Unit, CERN,
1211 Geneva 23, Switzerland;
Physics Department, Boston University, USA}%


\date{\today}

\begin{abstract}
A.M. Hillas, in a review of the origins of cosmic rays, has recently criticized
the ``cannonball" (CB) model of cosmic rays and gamma-ray bursts. We respond
to this critique and take the occasion to discuss the crucial question of particle
acceleration in the CB model and in the generally accepted models.
We also summarize our claims concerning the CB model.
\end{abstract}

\pacs{PACS Numbers}
\maketitle

\section{Motivation}
\label{s1}

The `cannonball' (CB) model is based on the hypothesis that a good fraction
of core-collapse SNe biaxially emit a few plasmoids of ordinary matter,
initially expanding (in their rest system) at the speed of sound in a 
relativistic plasma. The CBs have a typical initial Lorentz factor of
${\cal{O}}(10^3)$ and a baryon number of ${\cal{O}}(10^{50})$, roughly
corresponding to half the mass of Mercury. We contend that this model
provides a good, simple, few-parameter description of the properties of
cosmic rays (CRs), long-duration gamma-ray bursts (GRBs), X-ray flashes, 
the gamma background radiation, the natal kicks of neutron stars,
the magnetic fields of galaxy clusters, and the `cooling flows' of rich
clusters. 

The CB model is no doubt a considerable simplification, and may
of course be false. One way to express our claims ---undeniable,
in our opinion--- is the following. The model is an excellent mnemotechnic
pedagogical tool: practically every single property of the mentioned
phenomena can be derived, in the proverbial extent of the back of an envelope,
 from the stated hypothesis, a few auxiliary
observations (e.g.~the density of the interstellar medium) and elementary
physics considerations.

In an a-posteriori addition to a contribution to a conference \cite{Hillas},
A.M.~Hillas has made criticisms of the CB 
model. We are very grateful to receive an open
critique since, up to now, written comments on this model were confined
to the asymmetric environment of anonymous referee's reports.

In criticizing the CB model, Hillas gives a brief outline of it.
We will quote large portions of Hillas' article, since his 
outsider's explanations are also useful here, and are concordant
in style with his critical comments.
Quotes from Hillas will be in italics, e.g.~{\it The authors have gone 
on to estimate (most fully in Dar and de R\'{u}jula
\cite{DD2006}) that these objects} [CBs] {\it
would also naturally account for the generation of virtually
all cosmic rays as a consequence of  their motion through the interstellar gas.
However, I can give no credence 
to this model.}

\section{On the expansion of cannonballs}

{\it 
The vital properties of the cannonball are its Lorentz factor,
$\Gamma$, which determines the energy of the particles it emits, and its
transverse radius, $R$, which determines the rate at which particles are
swept up and the CB slows down.  
A quite false effect is employed 
in calculating that the expansion of $R$ is
quickly slowed down... }
{\it The swept-up particles, which have a
very high individual energy $\Gamma m_p c^2$ in the CB reference frame, 
and which are supposed to diffuse out of the CB, are said to exert an
inward pressure, opposing lateral expansion, supposedly because of momentum
reaction when they leave, whereas the opposite is true: near the edges, where
the net particle flow is outwards, a net {\it outward} force (pressure gradient)
would be exerted by the diffusing particles on the CB material.}

Consider the analogous problem of photons diffusing outwards in the
Sun and finally escaping freely from its ``surface": the photosphere. 
Their diffusion gradient indeed corresponds to a
volume-distributed upwards force. But the escaping photons do exert an
inwards force per unit surface on the photosphere. If Hillas were saying
the contrary, he would not be contradicting us, but Newton.
Our arguments concerning expansion apply to the
whole CB or, in the analogy, to the inwards pressure on the Sun's photosphere.

We explicitly admit in our papers to not having solved the problem
of the evolution of a CB from first principles. But we recall how
the relativistic
plasmoids ejected by quasars appear not to expand laterally as they travel
for hundreds of kiloparsecs, before they finally stop and blow up.
This is an observation, recognized as a mysterious fact, that we try to explain.
Hillas ignores this fact, as well as the success of our 
entirely analogous choice 
of CB-radius evolution in the description of the afterglows of GRBs.

\section{On the recapture of cosmic rays}

{\it The authors suppose that particles can simply leave the CB's surface,
without Fermi ``bouncing'', whereas Achterberg et al.~\cite{Achten}
show that particles entering a plasma advancing with ultrarelativistic speed,
and scattered back out of it, cannot escape: even a small
external magnetic field retards them sufficiently that they are recaptured,
and a Fermi acceleration process is set up that dominates the spectrum.
(If $B$ were as
small as the normal Galactic field $\sim{2}\rm{\mu G}$ they would still be
recaptured within the CB radius until their rigidity exceeded $\sim 10^{17}$ V.)}

The calculations that Hillas quotes are not  relevant. 
One reason is that
a CB is preceded by the collimated flux of CRs it has previously produced,
and this flux is largely sufficient to wipe out the pre-existing magnetic field.
An analogous phenomenon is the effect of the Sun's coronal mass ejections on 
interplanetary magnetic fields.

A CB of Lorentz factor $\Gamma$ intercepts the
particles at rest in
 the ISM (mainly protons, of number density $n_p$)
and, via the dominant `elastic' process, re-emits them in a forward
cone of characteristic angular width $\theta=1/\Gamma$, with a
typical Lorentz factor $\gamma\!\sim\!\Gamma^2$. Let the CB and CR velocities,
in $c=1$ units, be $\beta_{_{\rm CB}}\approx 1\!-\!1/(2\Gamma^2)$,
and $\beta_{_{\rm CR}}\approx 1\!-\!1/(2\gamma^2)$.
The number of ISM particles intercepted by the CB in an element
of travel distance, $dx=\beta_{_{\rm CB}}\,c\,dt$, is $\pi\,R^2\,n_p\,dx$.
They travel forwards in a layer of
fore-shortened depth 
$d\bar{x}=dx(\beta_{_{\rm CR}}/\beta_{_{\rm CB}}-1)
\approx dx/(2\Gamma^2)$. At a later time $t$, 
let $d=\beta_{_{\rm CR}}\,c\,t$ be the distance travelled by the CRs of
our elemental volume, bigger than the distance $\beta_{_{\rm CB}}\,c\,t$
travelled by the CB. 
The energy density of CRs inside the leading conical shell of radius
$\theta\,d$ and depth $dx'$ is:
\begin{eqnarray}
&&\epsilon_{_{\rm CR}}={2\,R^2\,n_p\,\Gamma^6\,m_p\,c^2\over d^2}
\sim 2\!\times\!10^7\,{{\rm eV}\over {\rm cm}^{3}}
\nonumber\\
&\times& \left[{\Gamma\over 10^3}\right]^6
\left[R\over 10^{14}\,{\rm cm}\right]^2
{n_p\over 10^{-3}\,{\rm cm}^{-3}} \left[{10\,{\rm kpc}\over d}\right]^2
\label{endens}
\end{eqnarray}
This number is very large, with respect to the energy density 
$\epsilon_{\rm B}\!\sim\! 0.16$ eV/cm$^3$ of a $B\!=\!2\,\mu$G field. 
At the beginning of a CB's trajectory, $d$ is small and the CR energy density
in front of it
is enormous. Even a tiny fraction of CRs evading recapture
would suffice to erase any previous magnetic field so that the
process underlying Eq.~(\ref{endens}) can start. It could be argued that we
should have adiabatically `compressed' the ambient magnetic field by the
ratio $dx'/dx$. This would not change the conclusion. Moreover, it is
no doubt incorrect. The ISM into which the CRs are emitted was previously
totally ionized by the GRB's $\gamma$ rays, becoming a highly conducting plasma.
In such a medium the compressed magnetic fields pointing in various
directions would efficiently `reconnect' and `annihilate'.
The CB-generated CRs travel in a domain wherein there
is no magnetic field, long enough for the (decelerating) CB not to
catch up with them.

\section{Acceleration within a CB}

{\it The CB authors do wish to use Fermi acceleration to get particles to
$10^{21}$ V, however, but seem to misinterpret the reports of Frederiksen 
et al. \cite{fredurelmagf,fredelec}, as showing that
relevant acceleration of ions will occur entirely behind the shock, so they 
considered motion entirely there.}

The expression {\it behind the shock} may not be the right one,
for the quoted papers contain `collisionless shocks' as part
of their titles, but their results
do not show anything like the development
of a shock.  

{\it Frederiksen et al.~did note, though, that the magnetic field in the
external medium was required for significant energy gain by ions.}

This is what the quoted authors actually say:
{``Frederiksen et al.~\cite{fredelec} reported evidence for particle acceleration
with electron $\gamma$'s up to $\sim100$, 
in experiments 
with an external magnetic field present in the upstream plasma.  This is indeed 
a more promising scenario for particle acceleration experiments, 
although in the experiments by Nishikawa et al.~\cite{Nishi}
results with an external magnetic field
were similar to those without."} The cited `experiments' do not seem to 
us to refer to ions in this context of `external' magnetic
fields. Also, the results are not decisive. Acceleration 
 playing a crucial role in theories
of CRs and GRBs, we discuss it some more.

\section{More on acceleration in the mergers of plasmas}

The required efficiency of the acceleration mechanisms 
invoked in the CB model is orders of magnitude smaller than
that required in the SN remnant (SNR) model of 
CR production.
To generate the CR fluxes up to the `knees' 
(predicted to occur at $E(A)=2\,A\,m_p\,c^2\,\Gamma_0^2$, with
$\Gamma_0\sim 10^3$) {\bf no} CR `acceleration' mechanism is invoked
in the CB model. These CRs originate in a single `elastic collision' of a 
CB and an ISM particle at rest. In contrast, the CRs of SNR models
must laboriously be accelerated by multiple passages through
a non-relativistic shock. In the CB model only the CRs above the knee
are accelerated in the relativistic merger of two plasmas. This fraction
of the observed CR flux is $10^{-12}$ to $10^{-13}$ of the total.
Numerical experiments may not be able to `measure' such a
small fraction.

Similar differences in the required acceleration
efficiency appear in the comparison of the CB model with
 the standard `fireball' model of GRBs and their afterglows.
 The $\gamma$-ray spectrum of a GRB is a broken power law
 changing fast from $E^{-\alpha}$ to $E^{-\beta}$ at a `peak' energy
 $E_{\rm p}$ (typical values, predicted by the CB model, and observed,
 are: $\alpha\sim1$, $\beta\sim2.1$, $(1\!+\!z)E_{\rm p}\!\sim\!250$ keV).
 The largely dominant number of $\gamma$ rays below $E_{\rm p}$
 is generated in the elastic Compton scattering of ambient photons
 by (non-accelerated) electrons comoving with the CB, in very close
 analogy with the generation of CRs below the knee. At optical
 frequencies, the afterglow of a GRB is also a broken power law
 whose indices and  `bend' frequency are correctly predicted by the 
 CB model. Only the very small fraction of the afterglow light
 lying above the bend frequency is emitted by accelerated electrons.
 
 The broken power-law spectra of CRs, GRBs and their afterglows,
as we have seen, have a common explanation, and do not require,
except above their `knees/peaks/bends'  a population of
accelerated ions or electrons.  Naturally, the CBs themselves must have been
accelerated somehow. The SNR models also require a `first acceleration':
that of the SN shell; the fireball models require the acceleration of
their thin conical shells of relativistic $e^+e^-$ pairs. 
None of these 
pre-accelerations is convincingly understood on fundamental grounds.

\section{Conclusions}

We have presented some of the reasons why the critique of the CB model by Hillas 
does not convince us. More significant than the criticisms
of Hillas and many a referee is the general immunity to the evidence in favour
of the model. Some examples of this evidence are:
\begin{itemize}
\item
A few parameters
(initial Lorentz factor, typical mass and radio self-absorption frequency
of a CB) and their distributions are extracted from the model's accurate 
description of
GRB afterglows (AGs). {\bf GRB afterglows are understood}
\cite{AGoptical,AGradio}.
\item
The AG description was used to
conclude that  {\bf long-duration GRBs are made by supernovae} \cite{AGoptical}
and to predict the date in which GRB-associated  SNe  would be
detectable
\cite{DDD2003a}. One example is 
SN2003dh, associated with GRB 030329, which changed the GRB community's
 views from considering the GRB/SN association a marginal item
(see, e.g.~\cite{Waxman}) to a crucial one (see e.g.~\cite{Woosley}). 
The non-relativistic ejecta of a SN close to its GRB axis are faster than average:
these conventional SNe {\bf appear to be `hypernovae'} \cite{Woosley}, but they are not.
\item
At small redshifts
 the CBs of a GRB ought to be directly observable as superluminal
sources of their radio AG \cite{GRB1}. One example is the two-pulse (two-CB)
GRB 030329 \cite{DDD2003a,DDD2004a}, whose double-source radio AG was indeed observed
\cite{TFBK,DDD2004b}: {\bf GRB cannonballs have been seen}.
\item
Given the information extracted from their AGs, all
 the properties of the single $\gamma$-ray pulses of GRBs are predicted in the CB
model, in agreement with observations \cite{DD2004}. {\bf Long GRBs are understood}.
\item
The X-ray AGs of GRBs are predicted
to have a specific complicated structure \cite{AGoptical}. This has 
recently been corroborated in impressive detail by SWIFT 
\cite{Nousek,DDD2006a}. {\bf The early evolution
of CBs is understood.}
\item
The  geometrical and radiation-beaming properties of a CB
are trivial:  those of an effectively point-like relativistic source.
Consequently,
many observable properties of a GRB (e.g.~peak energy and isotropic
energy and luminosity) are related to one another by predicted power laws
\cite{DD2004,DD2001},
in agreement with observations \cite{DD2004,DDD2006b}.  
GRB radiation is extremely collimated:
{\bf a good fraction of core-collapse SNe emit GRBs.}

\item
{\bf X-ray flashes are nothing but GRBs observed at larger angles}, as supported
by their properties \cite{DD2004,DDD2003c}, including their origin in SN explosions 
\cite{DDD2003c,Pian}.
\item
The spectral shapes and relative abundances of CRs up to their knees
are predicted, and in agreement with observations \cite{DD2006}. The predictions
are parameter-free and do not invoke any iterative acceleration mechanism.
{\bf CRs are understood up to the knees} and their understanding is trivial.
\item
 The tiny fraction of CR flux above the knees requires, in the
 CB model, an iterative acceleration akin to that required at
 all energies in the standard models of CRs and GRBs. With only
 one parameter for what this fraction is, the observed fluxes
 of CRs above the knees (and above the analogous energies in
 GRBs and their AGs) are understood. {\bf The entire spectra of CRs and GRBs
 are explained} with a single source: cannonballs.
 \item
 The CB model also offers straightforward expla-
nations of the natal kicks of neutron stars \cite{NatalK},
the properties of the gamma background radiation
\cite{GBR,DDD2006c}, of cooling flow clusters \cite{CDD}, and of the intensity
of magnetic fields in the intergalactic space, including
that in galaxy clusters \cite{DDmagnetic}.
{\bf The CB model is a proposed simple description
of all high-energy astrophysical phenomena 
generated by relativistic jets.}
\end{itemize}

Faced with these items, we conclude that there is
a choice between Okcam's razor and the opposite view:
{\it For every complex natural phenomenon there is a
simple, elegant, compelling, wrong explanation} (Tommy
Gold).
\\
\\

{\bf Acknowledgements.}
We are indebted to 
Andy Cohen, Shlomo Dado and
Shelly Glashow for useful discussions. This research was supported in part by the
Helen Asher Space Research Fund at the Technion Institute. 
\\


\begin{thebibliography}{99}

\bibitem{Hillas}
A.M. Hillas, astro-ph/0607109

 \bibitem{DD2006} 
 A. Dar  and A. De R\'{u}jula,
     hep-ph/0606199
     
 \bibitem{Achten}
A.  Achterberg et al.
 MNRAS  {\bf 328}, 393 (2001)
      
 \bibitem{fredurelmagf} 
 J.T. Frederiksen et al.,     Astrophys. J. {\bf 608}, L13 (2004)
 
 \bibitem{fredelec} 
 J.T. Frederiksen et al., astro-ph/0303360      
 
 \bibitem{Nishi}
K.-I. Nishikawa et al.,
Astrophys. J. {\bf 595}, 555 (2003)

\bibitem{AGoptical}
S. Dado, A. Dar and A. De R\'ujula, {Astron.\ Astrophys.\/} {\bf 388}, 1079 (2002)
 
\bibitem{AGradio}
S. Dado, A. Dar and A. De R\'ujula, { Astron.\ Astrophys.\/} {\bf 401},  243 (2003)

\bibitem{DDD2003a}

S. Dado, A. Dar and A. De R\'ujula, {Astrophys.\ J.\/} {\bf 594}, L89 (2003)

\bibitem{Waxman}
E. Waxman, Lect. Notes Phys. {\bf 598},  393 (2003)

\bibitem{Woosley}
    S. E. Woosley and J. S. Bloom,
    Annu. Rev. Astron. Astroph. {\bf 44}, 507 (2006)

\bibitem{GRB1}
A. Dar and A. De R\'ujula, astro-ph/0008474 

\bibitem{DDD2004a}
S. Dado, A. Dar and A. De R\'ujula,
astro-ph/0402374

\bibitem{TFBK}
G. B. Taylor, D. A. Frail, E. Berger and S. R. Kulkarni,
 Astrophys. J. {\bf 609},  L1 (2004)

\bibitem{DDD2004b}
S. Dado, A. Dar and A. De R\'ujula,
astro-ph/0406325

\bibitem{DD2004}
A. Dar and A. De R\'ujula, Phys. Rep. {\bf 405},  203 (2004)


\bibitem{Nousek}
J. A. Nousek et al., Astrophys. J. {\bf 642}, 389 (2006) 

\bibitem{DDD2006a}
S. Dado, A. Dar and A. De R\'ujula,
Astrophys. J. {\bf 646},  L21 (2006)


\bibitem{DD2001}
 A. Dar and A. De R\'ujula,
astro-ph/0012227

\bibitem{DDD2006b}
S. Dado, A. Dar and A. De R\'ujula,
astro-ph/0611161

\bibitem{DDD2003c}
S. Dado, A. Dar and A. De R\'ujula, { Astron.\ Astrophys.\/} {\bf 422}, 381 (2004)

\bibitem{Pian}
E. Pian et al., Nature {\bf 442},  1011 (2006)

\bibitem{NatalK}
A. Dar and R. Plaga,
Astron. Astrophys. {\bf 349},  259 (1999)

\bibitem{GBR}

A. Dar and A. De R\'ujula, MNRAS {\bf 323}, 391 (2001)

\bibitem{DDD2006c}
S. Dado, A. Dar and A. De R\'ujula
astro-ph/0607479

\bibitem{CDD}
 S. Colafrancesco, A. Dar and A. De R\'ujula,
Astron. Astrophys. {\bf 413},  441 (2004)
 
 \bibitem{DDmagnetic}
 A. Dar and A. De R\'ujula,
 Phys. Rev. {\bf D72},  123002 (2005)

\end{thebibliography}
\end{document}